\documentclass[final,5p,times,twocolumn]{elsarticle}

\usepackage{epsfig}

\usepackage{amssymb}

\journal{Optics Communications}

\begin{document}

\begin{frontmatter}



\title{Slow Light of an Amplitude Modulated Gaussian Pulse in Cesium Vapor}


\author{Wenzhuo Tang, Bin Luo, Yu Liu, and Hong Guo$^*$}

\address{CREAM Group, State Key Laboratory of Advanced Optical
Communication Systems and Networks (Peking University) and Institute of Quantum Electronics, School of Electronics Engineering and Computer Science (EECS), Peking University, Beijing 100871, China \\%
$^*$Corresponding author. Tel.: +86 10 62757035; fax: +86 10
62753208. E-mail address: hongguo@pku.edu.cn (H. Guo).}

\begin{abstract}Slow light of an amplitude modulated Gaussian (AMG) pulse in cesium
vapor is demonstrated and studied, as an appropriate amplitude
modulation to a single pulse can expand its spectrum and thus
increase the utilization efficiency of the bandwidth of a slow light
system. In a single-$\Lambda$ type electromagnetically induced
transparency (EIT) system, the slowed AMG pulse experiences severe
distortion, mainly owing to the frequency dependent transmission of
medium. Additionally, due to its spectral distribution, the
frequency dependent dispersion of the medium causes simultaneous
slow and fast light of different spectral components and thus a
certain dispersive distortion of the AMG pulse. Further, a
post-processing method is proposed to recover the slowed (distorted)
pulse, which indicates that by introducing a linear optical system
with a desired gain spectrum we can recover the pulse in an
``all-optical" way. Finally, we discuss the limitations during this
compensation procedure in detail. Although it is demonstrated in the
cesium vapor using EIT, this method should be applicable to a wide
range of slow light systems.
\end{abstract}

\begin{keyword}
Electromagnetically induced transparency (EIT) \sep slow light,
Gaussian pulse \sep distortion, cesium vapor \sep amplitude
modulation \sep delay line.

\PACS 42.50.Gy \sep 42.50.Ct \sep 42.30.Lr

\end{keyword}

\end{frontmatter}




\section{Introduction}
\label{introduction} Recently, slow light has attracted tremendous
attention for its potential applications in optical communication.
Group velocity of an optical pulse can be slowed down using various
effects, such as electromagnetically induced transparency (EIT)
\cite{Harris1997,F-RMP}, double absorption line
\cite{Howell1,Howell2}, coherent population oscillations
\cite{cpo1,cpo2,cpo3}, stimulated Brillouin scattering
\cite{sbs1,sbs2} and stimulated Raman scattering \cite{srs1}, etc.
Note that, most of the slow light experiments are performed using a
Gaussian pulse, while a few non-Gaussian (such as rectangularly
\cite{rect2005} and exponentially \cite{Sanghoon2009} shaped) pulses
are studied. These studies are based on the pulse shaping, while an
additional amplitude modulation to the pulse has not been well
investigated yet. Moreover, the study on the slow light of a pulse
with more complex spectrum should be very promising to the
applications. So, we believe that an amplitude modulated pulse has a
potential application in the optical communication. In this paper,
we demonstrate the slow light of a cosine-type AMG pulse
experimentally, and study both the absorptive and the dispersive
properties of this pulse in depth.

\begin{figure}[htb]
\centerline{\includegraphics[width=9.0cm]{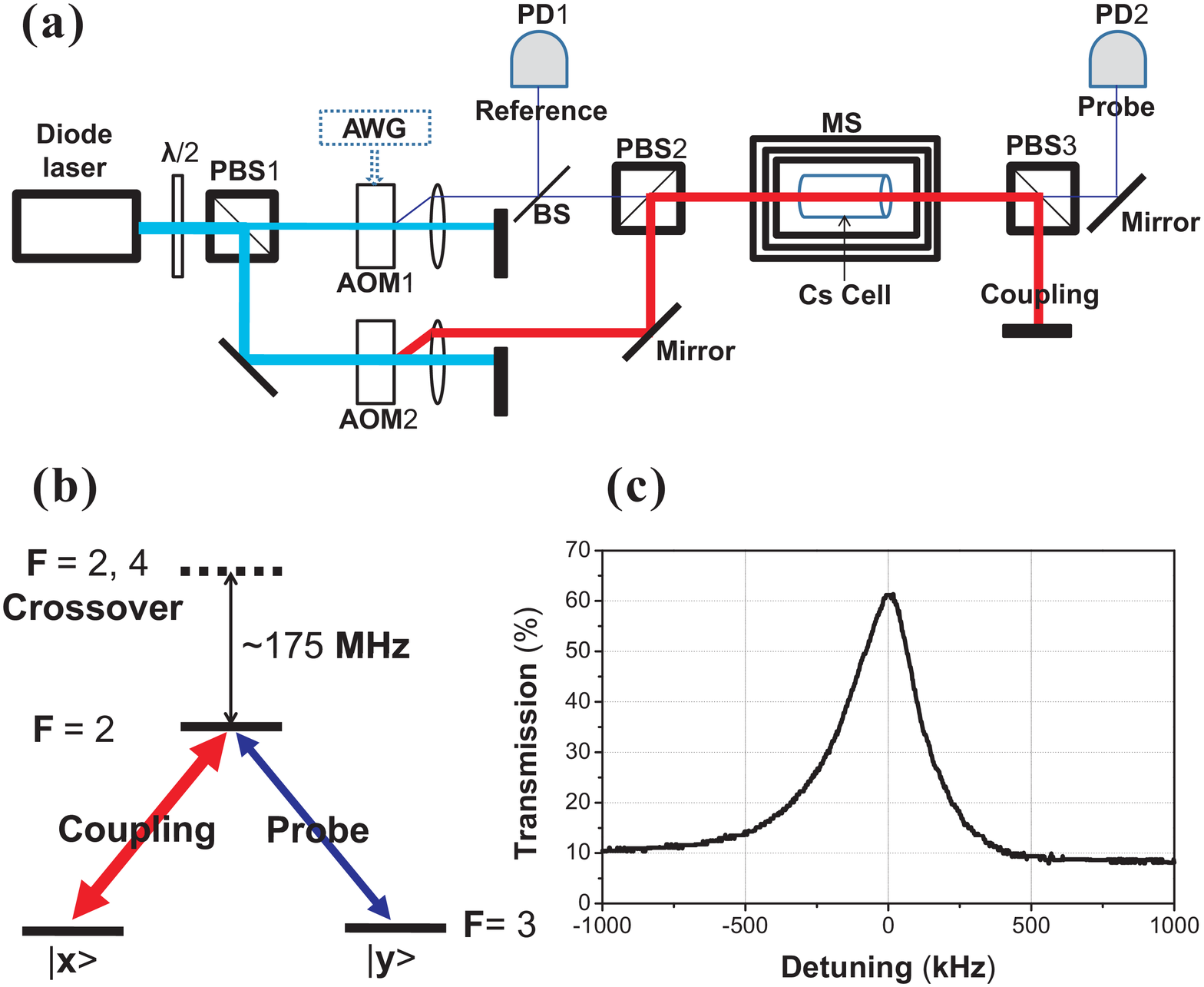}}
\caption{(color online). (a) Experimental setup. $\lambda /2$,
half-wave plate; PBS, polarizing beam splitter; AOM, acousto-optic
modulator; AWG, arbitrary waveform generator; MS, magnetic shield;
APD, avalanche photodetector. (b) A simplified single-$\Lambda$ type
system of \mbox{$^{133}$Cs} atom interacts with the coupling and the
probe lasers. Two ground states $|x\rangle$ and $|y\rangle$
represent appropriate superpositions of magnetic sublevels. (c)
Probe transmission spectrum (an average of 16 measured results)
versus detuning from the two-photon resonance, with the FWHM
bandwidth of \mbox{$\sim$350 kHz} and the maximum (background)
transmission of \mbox{$\sim$61.5\%}
(\mbox{$\sim$10\%}).}\label{SETUP}
\end{figure}

\section{Experiment}
\label{experiment} Our experimental setup is schematically
illustrated in Fig. \ref{SETUP}(a). An \mbox{852 nm} narrowband
(\mbox{$\sim$300 kHz}) diode laser (output power \mbox{$\sim$30 mW})
is stabilized to the $D_2$ transition of $^{133}$Cs atom: $F=3
\rightarrow F'=2, 4$ crossover. By the combination of $\lambda/2$
(half-wave plate) and PBS1, the laser is divided into two beams, the
coupling and the probe, with perpendicular polarizations and
adjustable proportions. The frequency of coupling laser is
redshifted \mbox{$\sim$175 MHz} by AOM2, with the power
\mbox{$\sim$2.35 mW} and $1/e^2$ beam diameter \mbox{$\sim$2.0 mm}.
The probe pulse is generated by passing a weak (\mbox{$<$10 $\mu$W})
continuous-wave laser through AOM1, with the frequency redshifted
\mbox{$\sim$175 MHz} and $1/e^2$ beam diameter \mbox{$\sim$2.0 mm}.
The amplitude of probe field $E(t)$ (nonnegative for convenience in
data processing) is proportional to the amplitude of the driving RF
(radio frequency) wave to AOM1, which is directly controlled by an
arbitrary waveform generator (AWG). After that, both the coupling
(vertically polarized) and the probe (horizontally polarized) lasers
are resonant with the $D_2$ transition $F=3 \rightarrow F'=2$ of
$^{133}$Cs atom [see Fig. \ref{SETUP}(b)], and combined at PBS2.
Then, the two overlapped lasers copropagate through the cell in
order to reduce the total Doppler broadening \cite{xiaomin1995}. The
beam splitter (BS) in front of the cell splits an appropriate
portion of the probe to APD1 as reference, whose intensity is set to
be equal to that of the output when the probe is detuned far
off-resonant from all atomic transitions. Thus, the background
absorption in experiment is eliminated and the reference can be
treated as ``input" in the following analysis. A \mbox{10 cm}-long
paraffin-coated cesium (\mbox{$^{133}$Cs}) vapor cell at room
temperature (\mbox{$\sim$25 $^{\circ}$C}) is placed in the magnetic
shield (MS) to screen out the earth magnetic field in lab. The
exiting beams are separated by PBS3 and only the slowed probe pulse
beam reaches APD2 as the ``output" in the following analysis.

In experiment, both the optical intensities $I(t)$ of the reference
(treated as ``input") and output pulses are recorded on a \mbox{100
MHz} digital oscilloscope triggered by the AWG. Thus, the amplitudes
of both input and output (slowed) pulses can be obtained from this
relation: $E(t)=\sqrt{I(t)}$, which is the reason of requiring the
amplitudes of probe pulses to be nonnegative in experiment. The
intensities of two probe (input) pulses in time domain are chosen as
\begin{eqnarray}
\mathrm{Gaussian:\
}I_1(t)&=&|E_1(t)|^2=\exp\left[\displaystyle-\frac{\displaystyle
(\ln2)t^2}{\displaystyle T_0^2}\right],\nonumber \\
\mathrm{AMG:\ }I_2(t)&=&|E_2(t)|^2=|E_1(t)\left[1+A\cos(2\pi{\delta}t)\right]|^2 \nonumber\\
&=&\exp\left[\displaystyle-\frac{\displaystyle(\ln2)t^2}{\displaystyle
T_0^2}\right]\left[1+A\cos(2\pi{\delta}t)\right]^2,\nonumber
\end{eqnarray}where $I_1(t)$ and $I_2(t)$ are the intensities of two pulses,
respectively; \mbox{$\delta$ (= 700 kHz)} is the modulation
frequency, $A$ (= 1) is the modulation depth, and $T_0$ ($\approx$
6.5 {\rm$\mu$}s) is the temporal duration (FWHM) of the intensity of
the Gaussian pulse.

Experimental results are shown in Fig. \ref{TIME}. The Gaussian and
the AMG pulses exhibit quite different slow light effects. In the
Gaussian case, the pulse [Fig. \ref{TIME}(a)] is delayed by
\mbox{$\sim$0.47 $\mu$s}, with relatively low loss and little pulse
distortion. Hence, we can compensate for the loss by directly
amplifying the intensity of the pulse with a constant magnitude
determined by the system. However, in the AMG case [Fig.
\ref{TIME}(b)], the output pulse undergoes relatively high loss and
significant distortion [meanwhile the delay time (\mbox{$\sim$0.14
$\mu$s}) is obviously decreased]. Thus, the direct amplification is
not feasible any more [see the ``rescaled" pulse in Fig.
\ref{TIME}(b)].

\section{Distortion due to absorption and dispersion}
\label{Distortion due to absorption and dispersion}

It is known that the pulse distortion is caused by both absorption
and dispersion \cite{Michael2005}. In an ideal EIT medium, when the
coupling field ($\Omega$) keeps unchanged both temporally and
spatially, the frequency response function of the slow light system
can be expressed as $H(\Delta)=\exp[-\Delta
Z/(\Delta-i\Gamma_{EIT})]$, where $\Gamma_{EIT}=|\Omega|^2/\gamma$
(proportional to the intensity of coupling laser) is the bandwidth
of EIT spectrum, $Z$ (proportional to the atomic number and cell
length) is the normalized propagation length, and $\Delta$ is the
detuning of probe laser. Suppose the EIT slow-light system is
linear, we have

\begin{eqnarray}
E_{out}(\Delta)&=&H(\Delta)E_{in}(\Delta)\nonumber \\
&=&\exp\left[\frac{\displaystyle-\Delta Z}{\displaystyle
 \Delta-i\Gamma_{EIT}}\right]E_{in}(\Delta) \nonumber \\
&=&\exp\left[\frac{\displaystyle -\Delta^2\gamma^2 Z}{\displaystyle
\Delta^2\gamma^2+|\Omega|^4}\right]\exp\left[\frac{\displaystyle -i\Delta \gamma Z |\Omega|^2}{\displaystyle \Delta^2 \gamma^2+|\Omega|^4}\right]E_{in}(\Delta) \nonumber \\
&\equiv&A(\Delta)e^{\displaystyle-i\Phi{(\Delta)}}E_{in}(\Delta),\nonumber
\end{eqnarray}where $E_{in}(\Delta)[E_{out}(\Delta)]$ represents the amplitude
spectrum of the input (output) pulse. Real functions
$A(\Delta)\equiv \exp\left[-\Delta^2\gamma^2
Z/(\Delta^2\gamma^2+\Omega^4)\right]$ [corresponding to the measured
transmission spectrum in Fig. \ref{SETUP}(c)] and
$\Phi(\Delta)\equiv \Delta \gamma Z
|\Omega|^2/(\displaystyle\Delta^2 \gamma^2+|\Omega|^4)$ are related
to the absorption and dispersion, respectively.

\subsection{Absorptive distortion}
\begin{figure}[htb]
\centerline{\includegraphics[width=9.0cm]{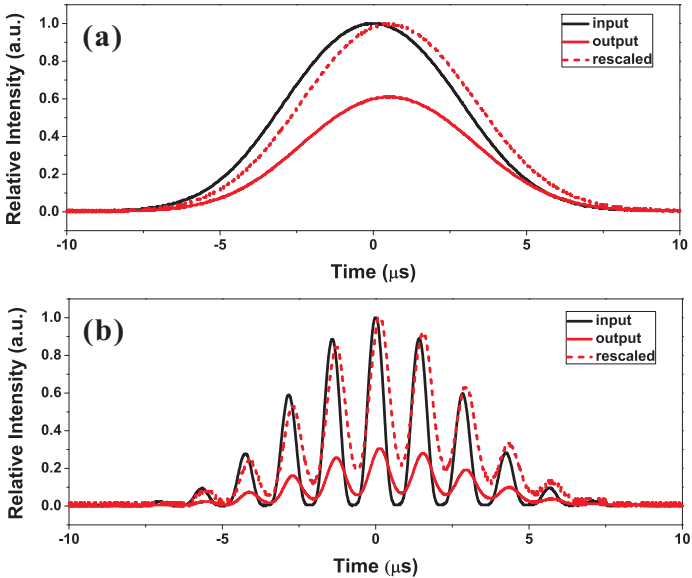}}
\caption{(color
online). Input (normalized) and output pulses in time domain. Each
pulse is an average of 16 measured results, and the ``rescaled" is
the normalization of the ``output". (a) Slow light of the Gaussian
pulse with low loss and little distortion. Pulse delay time
\mbox{$\sim$0.47 $\mu$s} is measured according to the peak of the
pulse. (b) Slow light of the cosine-type AMG pulse with high loss
and significant distortion. Pulse delay time \mbox{$\sim$0.14
$\mu$s} is measured according to the highest peak of the pulse.
}\label{TIME}
\end{figure}

\begin{figure}[htb]
\centerline{\includegraphics[width=9.0cm]{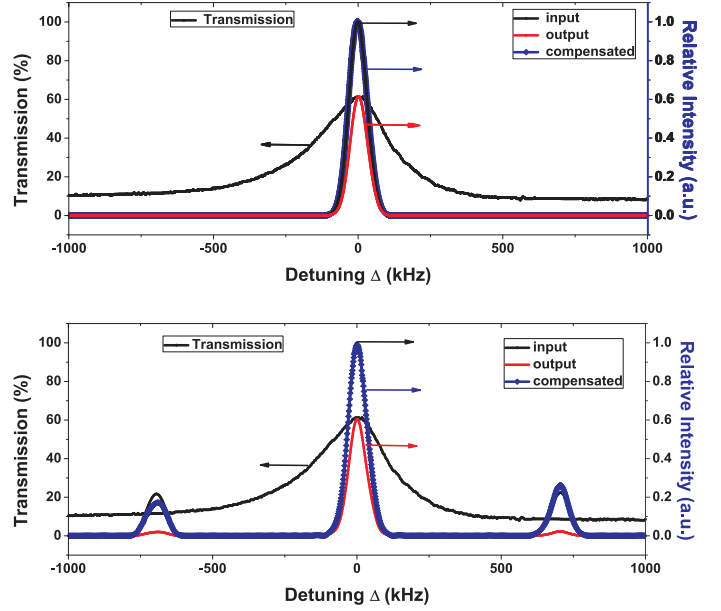}}
\caption{(color online). The intensity spectrums of the input (black
solid), output (red solid) and compensated (blue solid with bar)
pulses, and the transmission spectrum. (a) Gaussian case. The
compensated spectrum overlaps with the input one. (b) AMG case. The
compensated spectrum overlaps with the input one only with minor
deviations.}\label{FREQUENCY}
\end{figure}

In this section, we analyze the ``absorptive distortion" using the
transmission spectrum [see Fig. \ref{SETUP}(c)]. The intensity
spectrums of the input and output pulses are calculated out using
discrete Fourier transform (DFT) based on the experimental results.
It can be seen that the intensity spectrum of the Gaussian pulse
keeps unchanged with the FWHM bandwidth of \mbox{$\sim$74.4 kHz}
[Fig. \ref{FREQUENCY}(a)]; by contrast, the intensity spectrum of
the AMG pulse has not only a resonant Gaussian component, but also
two non-resonant Gaussian sidebands with \mbox{$\pm$700 kHz}
detunings symmetrically [Fig. \ref{FREQUENCY}(b)]. Since the
amplitudes of the optical field $E(t)$ and the spectrum $E(\Delta)$
satisfy the Fourier transform relation, this effect of generating
sidebands can be analyzed through the Fourier intensity spectrums of
the two input pulses:
\begin{eqnarray}
\mathrm{Gaussian:\ }I_1(\Delta)&=&|E_1(\Delta)|^2=\exp\left[\displaystyle-\frac{\displaystyle(\ln2)\Delta^2}{\displaystyle\Omega_0^2}\right],\nonumber \\
\mathrm{AMG:\ }I_2(\Delta)&=&|E_2(\Delta)|^2 \nonumber \\
&=&I_1(\Delta)+\frac{A^2}{4}\left[I_1(\Delta-\delta)+I_1(\Delta+\delta)\right],\nonumber
\end{eqnarray}where $I_1(\Delta)$ and $I_2(\Delta)$ are the intensity spectrums
of two pulses, respectively; ($A^2/4$) is the relative intensity of
the two sideband components; $\Omega_0$ (determined by $T_0$) is the
FWHM bandwidth of intensity spectrum of the Gaussian pulse. From the
spectrum [$I_2(\Delta)$] of the AMG pulse, we can see that the
``oscillation" of intensity in time domain [$I_2(t)$, in Fig.
\ref{TIME}(b)] is due to the interference of its three spectral
components: $I_1(\Delta)$, $I_1(\Delta-\delta)$, and
$I_1(\Delta+\delta)$.

Note that, we have $I_{out}(\Delta)=|A(\Delta)|^2\cdot
I_{in}(\Delta)$, where the phase information is eliminated, and
$|A(\Delta)|^2$ is just the transmission spectrum of the system [see
Fig. \ref{SETUP}(c)]. Three spectral components experience different
transmission rates according to the detuning ($\Delta$). In the AMG
case, two sideband components of the pulse undergo much more severe
absorptions than the resonant one. Thus, a compensation method is
proposed to recover the output slowed pulse. As shown in Fig.
\ref{FREQUENCY}, we obtain the compensated spectrums by simply
amplifying the output spectrums according to the transmission
spectrum, i.e.,
$I_{comp}(\Delta)=I_{out}(\Delta)/|A(\Delta)|^2_{measured}$. As the
compensated and the input spectrums are well overlapped in Fig.
\ref{FREQUENCY}, we can see that it is the frequency dependent
absorption that mainly causes the distortion of the AMG pulse owing
to the different loss of three spectral components.

\begin{figure}[htb]
\centerline{\includegraphics[width=9.0cm]{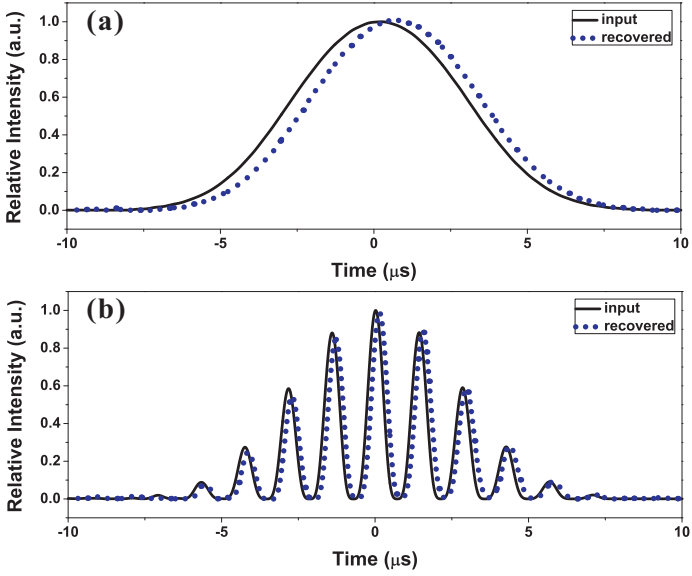}}
\caption{(color online). Input and recovered (calculated from the
compensated spectrum in Fig. \ref{FREQUENCY}) pulses in time domain.
(a) Gaussian case. Delay time \mbox{$\sim$0.46 $\mu$s} is measured
according to the peak of pulse. (b) AMG case. Delay time
\mbox{$\sim$0.11 $\mu$s} is measured according to the highest peak
of pulse. }\label{RECOVERED}
\end{figure}

It should be mentioned that, the noise is not considered in this
paper for its limited influence. Further, for a good compensation,
in experiment we use a scanning probe laser with slightly higher
intensity to obtain the transmission spectrum with a high SNR
(signal-noise-ratio), which also causes the $\sim$10\% ``background
transmission" as shown in Fig. \ref{SETUP}(c). Note that, on the
other hand, the probe laser (\mbox{$<$10 $\mu$W}) is weak enough
compared to the intensive coupling laser (\mbox{$\sim$2.35 mW}) and
its influence on the transmission spectrum can be ignored. To
further confirm the compensation result, we transform the
compensated spectrum back to time domain using the inverse discrete
Fourier transform (IDFT), and plot the input and the recovered
pulses in time domain together as shown in Fig. \ref{RECOVERED}. The
recovered pulses of Gaussian and AMG, with different time delays,
are both in good agreement with their corresponding input pulses in
shape and intensity. Especially in the AMG case, even with the
significant loss and distortion, both the intensity and the shape of
the slowed pulse are well recovered. In addition, although this
presented method is a post-processing one, a linear optical system
with the corresponding gain spectrum is able to achieve this
recovered result in an ``all-optical" way.

\begin{figure}[htb]
\centerline{\includegraphics[width=9.0cm]{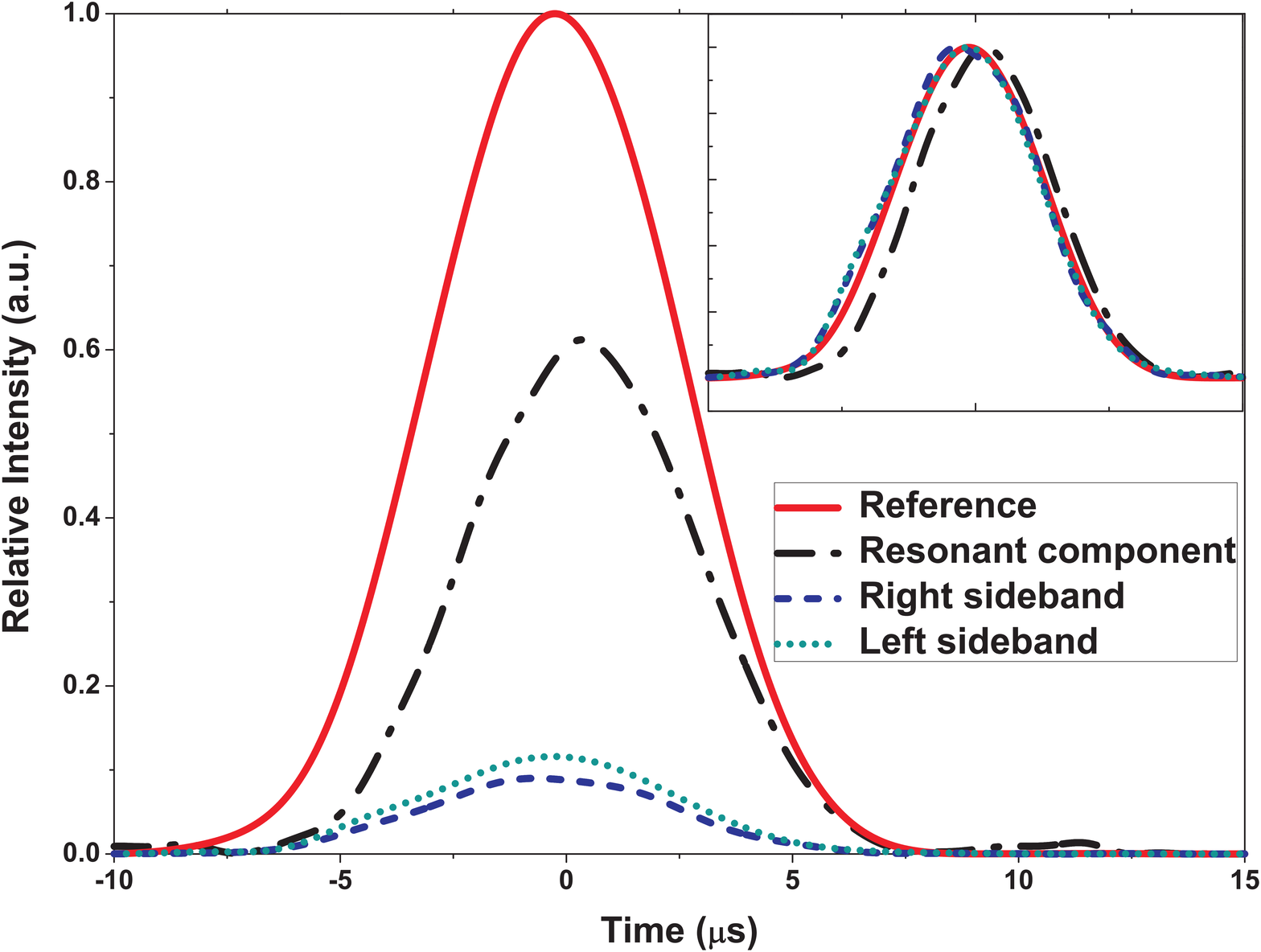}}
\caption{(color online). Simultaneous slow and fast light of the
separate spectral components of a single cosine-type AMG pulse.
Using IDFT, the resonant and two sideband components (left and
right) are obtained from the corresponding spectral components of
the output AMG pulse [Fig. \ref{FREQUENCY}(b)], while a common
reference pulse is obtained from the resonant spectral component of
the input AMG pulse. As a result, the resonant spectral component
propagates as slow light, while both the left and right sidebands
propagate as fast light. The delayed (advanced) time
\mbox{$\sim$0.47 $\mu$s} (\mbox{$\sim$0.18 $\mu$s} for left
sideband, \mbox{$\sim$0.20 $\mu$s} for right sideband) is measured
according to the peak of pulse. In the inset, all the pulses are
normalized for comparison.}\label{SFlight}
\end{figure}

\subsection{Dispersive dispersion}
\label{Dispersive distortion}

The different spectral components, according to $\Phi(\Delta)$, also
experience different phase shift, which leads to different
dispersion and introduces a certain distortion that is not severe in
our case. By separately transforming each spectral component back to
time domain using IDFT, we obtain three Gaussian pulses with
absolutely different ``time-delays". For the resonant component, we
obtain the ``slow light" of a Gaussian pulse [\mbox{$\sim$0.47
$\mu$s}, see Fig. \ref{SFlight}] with an identical time delay to the
non-modulated Gaussian one [\mbox{$\sim$0.47 $\mu$s}, see Fig.
\ref{RECOVERED}(a)]. On the contrary, for each sideband (left or
right) component, we obtain a Gaussian pulse even with a little time
advancing [\mbox{$\sim$0.18 $\mu$s} or \mbox{$\sim$0.20 $\mu$s}, see
Fig. \ref{SFlight}], namely ``fast light". It should be pointed out
that this phenomenon, called ``simultaneous slow and fast light", is
of different spectral components of a single pulse while that
presented in \cite{YifuZhu2006} is of two fields with the opposite
polarizations in the cold Rb atoms. The intensities of the left and
right sideband spectral components (fast light) in Fig.
\ref{SFlight} have a minor difference, since they experience
different absorptions due to the asymmetry of energy level structure
[see Fig. \ref{SETUP}(c)]. As a result, the time delay of the slowed
AMG pulse is decreased since it is a result of the interfering
superposition of these three separate spectral components, which
also tells why the Gaussian and AMG pulses experience different
time-delays [see Fig. \ref{TIME}].

On the other hand, this ``simultaneous slow and fast light" effect
in a single pulse will inevitably cause an additional pulse
distortion (``dispersive distortion"), but here it is negligibly
small for such a single-$\Lambda$ type system
\cite{Boyd2005,Camacho2006}. Actually, a good compensating result
has been obtained without eliminating the dispersive distortion
here. However, if the dispersive distortion is so severe that a
single AMG pulse splits into two or three pulses, the compensation
method we proposed is no longer valid. In addition, the modulation
frequency ($\delta$), determining the detuning and thus the
absorption and dispersion of the sideband components, have a
significant impact on the compensation result. Further, there is
always a residual dispersive distortion unless $\delta$ is small
enough that all the spectral components are within the linear
dispersion region of EIT spectrum. As another influencing factor,
the spectral bandwidth [$\Omega_0$, inversely proportional to the
temporal FWHM ($T_0$)] of three components has a relatively small
impact to the compensation, as long as $\Omega_0$ ($T_0$) is not too
large (small). Finally, by using the totally opposite dispersive
regions, a single EIT medium can simultaneously generate slow and
fast light for a single cosine-type AMG pulse with the suitable
modulation frequency. In this way, this amplitude modulation
increases the utilization efficiency of the bandwidth of the
EIT-based slow light system.

Further, the delay time of the Gaussian pulse [\mbox{$\sim$0.47
$\mu$s}, see Fig. \ref{TIME}(a)] remains approximately unchanged
after compensation [\mbox{$\sim$0.46 $\mu$s}, see Fig.
\ref{RECOVERED}(a)], because its spectrum has relatively intensive
distribution within the transmission spectrum. By contrast, the
delay time of the recovered AMG pulse [\mbox{$\sim$0.11 $\mu$s}, see
Fig. \ref{RECOVERED}(b)] is decreased after the compensating process
[compared with the delay \mbox{$\sim$0.14 $\mu$s of the output pulse
in Fig. \ref{TIME}(b)}]. That is to say, to a certain extent, the
compensation method weakens the slow light effect of the AMG pulse.
This is because that, these separate spectral components propagating
with definitely different group velocities (in our case, slow and
fast light), are amplified by different magnitudes during the
compensating process. The fast light effect of the two sideband
components are enlarged more than the slow light effect of the
resonant one, so that the delay time of the recovered AMG pulse,
which is the interfering superposition of the three spectral
components, is undoubtedly reduced. In addition, obviously, a larger
modulation depth ($A$) will lead to the larger delay reduction and
more significant residual distortion (caused by frequency dependent
dispersion) after compensation. Therefore, when applying this
compensation, two aspects need to be considered. Firstly, we should
evaluate the impact of three parameters of the amplitude modulated
pulse, i.e., modulation frequency ($\delta$), temporal FWHM ($T_0$)
and modulation depth ($A$), and then determine whether a
well-recovered (the residual distortion is small enough) pulse can
be obtained. Secondly, we need to make a trade-off between the pulse
shape (intensity) and time delay, i.e., to determine whether it is
worth while to compensate the loss and distortion with such a
reduction of time delay.

\section{Conclusion}
\label{conclusion}

\label{conclusion}To our knowledge, the slow light of a cosine-type
AMG pulse, which can be used to increase the utilization efficiency
of the bandwidth of slow light system by using the different
dispersion regions of the EIT spectrum, is demonstrated and studied
for the first time. Taking the cosine-type AMG pulse as an example,
we focus on studying the significant pulse distortion (loss) in a
single-$\Lambda$ EIT system. Then we point out that the distortion
is mainly caused by the frequency dependent transmission.
Accordingly, we present a compensation method to recover the slowed
pulses in shape and intensity with high fidelity, and the results
also indicate that, by introducing a linear optical system with
desired gain spectrum, the pulse can be recovered in an
``all-optical" way. On the other hand, due to the spectral
distribution of the cosine-type AMG pulse, the frequency dependent
dispersion causes an (minor) additional distortion and a reduction
of time delay. Meanwhile, distinguished from \cite{YifuZhu2006}, we
reveals the ``simultaneous slow and fast light" effect of different
spectral components in a single amplitude modulated optical pulse.
Further, according to the absorptive and dispersive properties of
slow light medium, the group delay and shape distortion of a pulse
can be manipulated respectively, while the absorptive distortion is
well compensated using the method presented in this paper. In
principle, this method is applicable to any other amplitude
modulated pulses, and is also extendable to many other slow light
systems. It is reasonable to believe that the methods of an
additional amplitude modulation to a single pulse and recovery of
slowed (distorted) pulse in slow light would have potential
applications in the optical communication and the low-distortion
optical delay lines.

\section*{Acknowledgement}

The authors appreciate A. Dang, S. Yu, C. Zhou, S. Gao, and W. Wei
for the helpful discussions. This work is supported by the Key
Project of the National Natural Science Foundation of China (Grant
No. 60837004), and the Open Fund of Key Laboratory of Optical
Communication and Lightwave Technologies (Beijing University of
Posts and Telecommunications), Ministry of Education, P. R. China.

\end{document}